\documentstyle[11pt]{article}
\begin{document}
\centerline{\Large\bf Singularities in kinetic theory}
\vskip 30pt
\centerline{C.Y. Chen}
\centerline{Dept. of Physics, Beijing University of Aeronautics}
\centerline{and Astronautics, Beijing 100083, PRC}

\vskip 20pt
\centerline{Email: cychen@public2.east.net.cn}

\vfill
{
\noindent {\bf Abstract:}
It is revealed that distribution functions of practical gases relate to 
singularities and such singularities can, with molecular motion, spread to 
the entire region of interest. It is also shown that even common continuous 
distribution functions involve a similar quasi-discontinuity difficulty.

}
\newpage
It is widely assumed that the necessary and complete foundation of kinetic 
theory has been established and the main problem in this field is the lack 
of powerful computational means: if a sufficiently large supercomputer were 
available we would be able to solve the well-known Boltzmann-type equations 
in terms of initial and boundary conditions and the behavior of most 
practical gases would become analyzable and 
calculable\cite{reif}\cite{wirz}. In view of that 
similar partial differential equations in fluid mechanics can be handled by 
today's computers and the obtained results appear to be reasonably good (to a 
certain extent), it seems that the aforementioned notion concerning 
kinetic theory is indeed sound and cannot, thus should not,
be challenged seriously.

In studying several kinds of gases one, 
however, found that singularities arose from the standard treatment\cite{chen0}. 
Unlike singularities that exist for usual partial differential systems, the 
observed singularities are not limited to isolated regions; molecular motion 
carries them to nearby regions as well as distant regions 
(though collisions between molecules erase them partly).
 Being exposed to such difficulties, we no longer believe that the principal way of 
studying kinetic gases is to have Boltzmann-type equations.

The purpose of this paper is to arouse concern for the singularity 
aspect of the standard kinetic theory.  We present realistic examples in 
that discontinuous distribution functions come to exist and application of differential
operator to them becomes risky and misleading. Most of analyses 
are carried out in terms of solving the collisionless Boltzmann 
equation numerically and analytically. At the end we briefly discuss 
in what way we may overcome the difficulties revealed herein. 

First of all, let us take a brief look at how the Boltzmann equation 
can be solved in terms of the standard procedure. The equation reads
\begin{equation}\label{bl} \frac{\partial f}{\partial t}+{\bf v}\cdot 
\frac{\partial f}{\partial {\bf r}} 
+\frac {{\bf F}}m \cdot\frac{\partial f}{\partial 
{\bf v}}=\int[f({\bf v}^\prime)f({\bf v}_1^\prime)-f({\bf v})f({\bf 
v}_1)]u\sigma d\Omega d{\bf v}_1 , \end{equation}
where the left side is sometimes called the convective operator and the 
right side the collisional operator. The detailed meaning of the notations 
related to two operators can be found in any 
textbook\cite{reif}\cite{harris}. To solve the differential-integral 
equation, we must have the initial condition
\begin{equation} f(t_0+0,{\bf r},{\bf v})=f(t_0,{\bf r},{\bf v}) 
\end{equation} and the boundary condition
\begin{equation}\label{bc}
 f(t,{\bf r},{\bf v})=\int K({\bf v},{\bf v}_1) f(t,{\bf r},{\bf v}_1)d{\bf 
v}_1, \end{equation}
where $K({\bf v},{\bf v}_1)$ represents the collisional nucleus between 
molecules and boundaries\cite{wirz}\cite{kogan}. According to the existing 
kinetic theory, these equations, from (\ref{bl}) to (\ref{bc}), constitute a 
complete equation set to describe the gas dynamics. It is also noted that 
to numerically solve the equation set a finite-difference 
scheme based on a grid arrangement in the position-velocity space, such as  
\begin{equation} \frac{\partial f}{\partial t}\sim \frac{\Delta f}{\Delta 
t},\quad \frac{\partial f}{\partial x}\sim \frac{\Delta f}{\Delta x},\quad
\frac{\partial f}{\partial v_x}\sim \frac{\Delta f}{\Delta 
v_x},\cdots\cdots,\end{equation} 
should be employed. If all these are ready, we can, by means of a ``true''
supercomputer, construct solutions for the equation set. 
For convenience of later discussion, we will refer to the procedure outlined above, 
including the equations and all the necessary mathematical treatments, as {\it the 
standard solution-construction scheme} of the Boltzmann equation. 

Enormous difficulties involve in the solution-construction scheme. What 
have been well known include: (i) There are seven independent variables: 
time, geometric coordinates, and velocity components. (ii) The collisional 
operator is complicated in nature. Due to these difficulties, only much 
simplified cases have been investigated (one-dimensional cases for 
instance). In this paper, we put aside these well-known difficulties and try 
to discuss what may happen when a sufficiently powerful supercomputer is 
available and a full-dimensional computational job is really performed.

For simplicity, we will disregard collisions between molecules. 
Nevertheless, the discussion in this paper should hold its significance for rarefied 
gases as well as for ordinary gases.

Consider a boundary in the gas shown in Fig. 1. Suppose that there initially
exists a density difference between the two sides of the boundary 
and that the density on the left side is much larger. Obviously, molecules 
on the left side will expand into the space on the
right side. At any point on the right side, we will surely find out a 
certain amount of molecules that come from  the left side. By referring to 
Fig. 2a, we set up a local coordinate system at a point $p$ on the right 
side in such a way that the $y$-direction is parallel to the boundary edge 
and molecules having passed by the edge will move in the $x$-direction at 
the point. It is then simple to see that any molecules that have velocities 
$v_z<0$ do not come from the left side of the boundary. In Fig. 2b, we 
schematically plot the velocity distribution at $p$, which 
illustrates that the distribution function at $v_z=0$ involves a 
discontinuity \begin{equation}\label{fin} \frac{\partial f}{\partial 
v_z}=\infty.\end{equation} If the $z$-component of the external force is not 
exactly zero (there is a 
gravitational force, for instance) the convective operator in the Boltzmann 
equation fails to make sense at $v_z=0$
\begin{equation}\label{fin1}  
\frac{\partial f}{\partial t}+{\bf v}\cdot\frac{\partial f}{\partial {\bf 
r}}+ \frac{{\bf F}}m\cdot \frac{\partial f}{\partial {\bf v}}=\infty. 
\end{equation} The situation is rather worrisome since every spatial point
on the right side (as well as the left side) of the boundary involves exactly 
the same singularity for exactly the same reason. 

The singularity discussed above has some thing to do with external
forces. To have a complete picture, we will, in the remaining part of this 
paper, investigate examples in that no external force exists.

We consider a gas leaking out of a container through a small hole, as shown in Fig. 3, 
which schematically represents how a point-size source emits molecules and how the 
emitted molecules spread out over a free space.

At this point, it is worth mentioning that the situation under consideration 
is not particularly special. In addition to what happen to leaking gases, 
there are many practical cases in that molecules behave like an outflow from 
a point-like source. As shown in Fig. 4, a boundary surface cannot be 
regarded as a uniform one owning to physical, 
chemical and geometric differences on it;
 and this leaves us no choice but to divide the surface into many small, 
infinitesimally small according to the spirit of calculus, area elements and 
examine how those elements reflect molecules. Due to the finite temperature 
of the boundary, the ``reflected'' molecules from each of the area element
 will spread as if they are 
emitted from a small hole. (Emission patterns have been examined by Ref. 5, though from a 
somewhat different viewpoint.)

We first apply the ordinary fluid mechanics to the fluid around the point 
$p$ in Fig. 3b. If we further assume that the gas container is rather large, 
the outflow must be time-independent during the interested period, which 
means \begin{equation} \frac {\partial n}{\partial t}=0.\end{equation}
It is easy to find that at the point $p$
\begin{equation}\label{fluid1} \frac{\partial n}{\partial x}\not= 0,\quad 
\frac{\partial v_x}{\partial x}=0.\end{equation}
These expressions tell us that the usual fluid equation
\begin{equation}\label{neq} \frac{\partial n}{\partial t}+ \nabla ({\bf v} 
n) = 0\end{equation}
does not hold.

Can the standard kinetic theory do better? The immediate concern is to 
determine the distribution function related to the gas. By referring to  
Fig. 3b, in which the origin of the coordinate system is placed at the 
point-like source, we find that the distribution function can be expressed 
by 
\begin{equation}\label{outdis} f({\bf r},v,\Omega)=\frac{u(v) g(\Omega)}
{r^2}\delta(\Omega-\Omega_{{\bf r}}),\end{equation}
where $\Omega$ is the solid angle of the velocity and $\Omega_{{\bf r}}$ is 
the solid angle of ${\bf r}$ in the position space, $u(v)$ stands for a 
function of $v=|{\bf v}|$, which may, for instance, be proportional to 
$\exp(-\mu v^2/2)$ and $g(\Omega)$ represents a function of $\Omega$. 
In Eq. (\ref{outdis}) the factor $r^{-2}$ is due to the 
expansion of the molecular paths. It is rather obvious that expression 
(\ref{outdis}) can generally stand for an outflow emitted by a point-size 
source. In terms of this distribution function, we surely have, at the point 
$p$ again,
\begin{equation} \frac{\partial f}{\partial t}=0, \quad {\bf F}\cdot 
\frac{\partial f}{\partial {\bf v}}=0\end{equation}
and
\begin{equation} {\bf v}\cdot \frac{\partial f}{\partial {\bf r}}\not= 
0.\end{equation} Similar to what happens to the fluid equation (\ref{neq}), the 
collisionless Boltzmann equation 
\begin{equation}\label{bl0} \frac{\partial f}{\partial t}+{\bf v}\cdot 
\frac{\partial f}{\partial {\bf r}} 
+\frac {{\bf F}}m \cdot\frac{\partial f}{\partial 
{\bf v}}=0\end{equation}
is not valid.

When dealing with partial differential equations, it is customary to think 
of singularity as something isolated in a 
certain domain. The singularities revealed in this paper, however, are  
different: the point $p$ in Fig. 3, as well as the point $p$ in Fig. 2, is 
chosen rather arbitrarily, and this means such singularities exist in the 
entire space of interest.

To see the deep root of the difficulty, we wish to continue our discussion on
one unusual behavior of continuous distribution 
function. The behavior is well associated with the singularities
 that have just been discussed. 

In Fig. 5, we schematically depict a gas. Suppose that in the shaded region 
the density of molecules is significantly larger than those in the nearby 
regions (however, the continuity of the distribution function is still there). 
Instead of using the standard solution-construction scheme, we try a slightly different, 
but analytically much more effective, approach. Think about 
how the distribution function at the point $o$ influences the distribution 
function at $p_1$ and at $p_2$. By relating ${\bf r}$ and ${\bf v}$ in the
the collisionless Boltzmann equation to those of a moving molecule, we can
write the equation as  
\begin{equation}\label{path}\left.\frac{df}{dt}\right|_{\rm 
path}=0,\end{equation} where the ``path'' implies that the differentiation 
is taken along
a molecular path in the position-velocity phase space. The solution of Eq. 
(\ref{path}) is simply
\begin{equation}\label{path1} f(t,{\bf r},{\bf v})|_{\rm path}={\rm 
Constant}.\end{equation} 
In other words, we can link the distribution function at a certain point
to the distribution function at another point if the link exists in terms
of a molecule's path. For the situation shown in Fig. 5a, we know that
\begin{equation}\label{5a}
f(t_0,{\bf r}_o,{\bf v})= f(t_1,{\bf r}_{p_1},{\bf v})= f(t_2,{\bf r}_{p_2},
{\bf v}),\end{equation}
where ${\bf v}$ is the velocity of the moving molecule (no external force exists). 
In terms of (\ref{5a}), we
may say that $o$ is the ``source'' point and $p_1$ or $p_2$ is the 
``image'' point. 

The formulation above seems ``exactly'' consistent with the standard approach. 
However, there are several things worth special mentioning. As 
one thing, the path-information of molecules plays an active and essential 
role in this approach while it is considered almost irrelevant in the 
equation set (\ref{bl})-(\ref{bc}). As another 
thing, this approach is less sensitive to singularities. Equation (\ref{path}) 
is an ordinary differential equation along a path and the singularities 
associated with (\ref{fin}) and (\ref{fin1}), for instance,
 do not spell much trouble to it.     

The resultant expression (\ref{path}) or (\ref{path1}) brings out that, by 
referring to the Fig. 5b, the local distribution functions at $p_1$ 
and $p_2$ have cone-like structures. The structures are  
interesting in the following two senses. One is that the cones become sharper and 
sharper constantly as the distance between the source  and the 
image increases. The other is that though the initial variation of the
distribution function is in the spatial space, the cone-like structures are
formed later on in the velocity spaces of other points.

In Fig. 6, we plot the distribution function versus the polar
angle in the velocity space. The figure clearly illustrates that with the 
increase of the distance between the source and the image
\begin{equation}\label{quasi}\frac{\partial f}{\partial \theta}\rightarrow 
{\rm very\; large.}\end{equation}
If the expression
\begin{equation}\label{true}\frac{\partial f}{\partial \theta}\rightarrow 
\infty\end{equation} 
is allowed to characterize a true discontinuous distribution function, see for 
instance expression (\ref{outdis}), it should be appropriate to name the 
feature related to (\ref{quasi}) as the {\it quasi-discontinuity}.

It is now in order to comment on the applicability of the standard 
solution-construction scheme outlined at the beginning of the discussion. Equation 
(\ref{quasi}) has shown that even if we assume that the 
distribution function under consideration is initially continuous and the 
gas is free from boundary effects, the standard scheme will still encounter 
difficulties. As the cones of the velocity spaces become sharper and sharper, too 
sharp to be described by the chosen grid arrangement, 
some kinds of $\delta$-functions have to be employed. In this sense, no 
approach is truly usable unless a way is included in which both continuous 
and discontinuous distribution functions are treated on a roughly equal 
footing. 

Finally, we make a brief examination of possible direction in which the 
revealed difficulties can be surmounted. 

The discussions in this paper have shown that the most essential task is to 
deal with continuous and discontinuous distribution functions in a unifying 
way. After many unsuccessful tries, we are convinced that the task can be 
accomplished by an integral procedure in that the path-information of 
molecules plays an important role. The reasons for that include: (i) 
Integral operations, unlike differential operations, are usually not 
sensitive to discontinuity; if formulated adequately, the behavior of both 
discontinuous and continuous distribution functions can be described. (ii) 
In discussing the discontinuity and the quasi-discontinuity, we have seen 
that if the path-information of molecules is made of use, the mental 
picture, as well as the resultant formulas, becomes much clarified. 

Based on the conceptions aforementioned, we have developed a path-integral 
approach\cite{chen0}\cite{chen} in that 
the singularity difficulties revealed in this
paper are removed. Best of all, some of full-dimensional practical gases 
become calculable in terms of today's computers.

Discussion with Professor Keying Guan is gratefully acknowledged. His 
mathematical viewpoint on turbulence is one of the stimulating factors of 
this paper. The work is partly supported by the fund provided by Education 
Ministry, PRC.

\newpage

\centerline{\bf\Large Figure captions}

\begin{enumerate}
\item
A gas in that density difference is maintained by boundary blocking.
\item Discontinuous distribution function associated with boundary blocking.
(a) The local 
coordinate frame at a point $p$. (b) The distribution function versus $v_z$. 
\item A gas leaking out of a container through a small hole.
\item Schematic of molecules reflected by a boundary.
\item A dense gas influencing the nearby and distant regions. 
\item Distribution function in terms of the polar angle. (a) In a nearby 
region. (b) In a distant region. 

\end{enumerate}

\newpage

\noindent {\bf Figure 1}

\setlength{\unitlength}{0.013in} 
\begin{picture}(200,150)
\put(90,15){\makebox(35,8)[l]{\bf (a)}}
\put(270,15){\makebox(35,8)[l]{\bf (b)}}
\multiput(105,86)(180,0){2}{\line(0,-1){45}}
\multiput(106,86)(180,0){2}{\line(0,-1){45}}
\multiput(81,74)(0,15){4}{\vector(1,0){5}}
\multiput(260,59)(0,15){4}{\vector(2,1){5}}
\multiput(235.29,85.58)(0,-15){4}{\circle*{1}}
\multiput(237.96,87.85)(0,-15){4}{\circle*{1}}
\multiput(240.69,90.03)(0,-15){4}{\circle*{1}}
\multiput(243.47,92.17)(0,-15){4}{\circle*{1}}
\multiput(246.28,94.25)(0,-15){4}{\circle*{1}}
\multiput(249.09,96.33)(0,-15){4}{\circle*{1}}
\multiput(251.97,98.32)(0,-15){4}{\circle*{1}}
\multiput(254.90,100.24)(0,-15){4}{\circle*{1}}
\multiput(257.82,102.16)(0,-15){4}{\circle*{1}}
\multiput(260.82,103.97)(0,-15){4}{\circle*{1}}
\multiput(263.86,105.70)(0,-15){4}{\circle*{1}}
\multiput(266.95,107.36)(0,-15){4}{\circle*{1}}
\multiput(270.07,108.94)(0,-15){4}{\circle*{1}}
\multiput(273.20,110.51)(0,-15){4}{\circle*{1}}
\multiput(276.39,111.95)(0,-15){4}{\circle*{1}}
\multiput(279.62,113.29)(0,-15){4}{\circle*{1}}
\multiput(282.89,114.54)(0,-15){4}{\circle*{1}}
\multiput(286.20,115.68)(0,-15){3}{\circle*{1}}
\multiput(289.51,116.83)(0,-15){3}{\circle*{1}}
\multiput(292.87,117.80)(0,-15){3}{\circle*{1}}
\multiput(296.26,118.65)(0,-15){3}{\circle*{1}}
\multiput(299.69,119.38)(0,-15){3}{\circle*{1}}
\multiput(303.13,120.00)(0,-15){3}{\circle*{1}}
\multiput(306.59,120.51)(0,-15){3}{\circle*{1}}
\multiput(310.07,120.89)(0,-15){3}{\circle*{1}}
\multiput(313.55,121.27)(0,-15){3}{\circle*{1}}

\multiput(60.00,117.00)(0,-15){4}{\circle*{1}}
\multiput(63.44,117.62)(0,-15){4}{\circle*{1}}
\multiput(66.91,118.12)(0,-15){4}{\circle*{1}}
\multiput(70.39,118.50)(0,-15){4}{\circle*{1}}
\multiput(73.88,118.76)(0,-15){4}{\circle*{1}}
\multiput(77.38,118.90)(0,-15){4}{\circle*{1}}
\multiput(80.88,118.91)(0,-15){4}{\circle*{1}}
\multiput(84.37,118.80)(0,-15){4}{\circle*{1}}
\multiput(87.87,118.68)(0,-15){4}{\circle*{1}}
\multiput(91.36,118.34)(0,-15){4}{\circle*{1}}
\multiput(94.82,117.88)(0,-15){4}{\circle*{1}}
\multiput(98.27,117.29)(0,-15){4}{\circle*{1}}
\multiput(101.70,116.59)(0,-15){4}{\circle*{1}}
\multiput(105.11,115.77)(0,-15){3}{\circle*{1}}
\multiput(108.48,114.85)(0,-15){3}{\circle*{1}}
\multiput(111.86,113.92)(0,-15){3}{\circle*{1}}
\multiput(115.17,112.80)(0,-15){3}{\circle*{1}}
\multiput(118.45,111.58)(0,-15){3}{\circle*{1}}
\multiput(121.69,110.26)(0,-15){3}{\circle*{1}}
\multiput(124.90,108.85)(0,-15){3}{\circle*{1}}
\multiput(128.06,107.35)(0,-15){3}{\circle*{1}}
\multiput(131.22,105.85)(0,-15){3}{\circle*{1}}
\multiput(134.32,104.21)(0,-15){3}{\circle*{1}}
\multiput(137.36,102.49)(0,-15){3}{\circle*{1}}
\multiput(140.37,100.69)(0,-15){3}{\circle*{1}}
\multiput(143.37,98.90)(0,-15){3}{\circle*{1}}
\end{picture}
\vskip 15pt

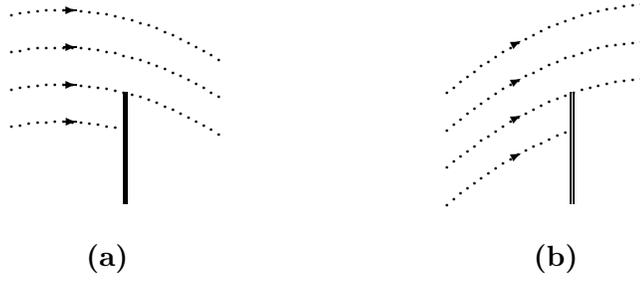
\noindent {\bf Figure 2}

\setlength{\unitlength}{0.013in} 
\begin{picture}(175,155)
\put(90,15){\makebox(35,8)[l]{\bf (a)}}
\put(270,15){\makebox(35,8)[l]{\bf (b)}}
\put(105,86){\line(0,-1){45}}
\put(106,86){\line(0,-1){45}}
\multiput(81,74)(0,15){4}{\vector(1,0){5}}
\multiput(60.00,117.00)(0,-15){4}{\circle*{1}}
\multiput(63.44,117.62)(0,-15){4}{\circle*{1}}
\multiput(66.91,118.12)(0,-15){4}{\circle*{1}}
\multiput(70.39,118.50)(0,-15){4}{\circle*{1}}
\multiput(73.88,118.76)(0,-15){4}{\circle*{1}}
\multiput(77.38,118.90)(0,-15){4}{\circle*{1}}
\multiput(80.88,118.91)(0,-15){4}{\circle*{1}}
\multiput(84.37,118.80)(0,-15){4}{\circle*{1}}
\multiput(87.87,118.68)(0,-15){4}{\circle*{1}}
\multiput(91.36,118.34)(0,-15){4}{\circle*{1}}
\multiput(94.82,117.88)(0,-15){4}{\circle*{1}}
\multiput(98.27,117.29)(0,-15){4}{\circle*{1}}
\multiput(101.70,116.59)(0,-15){4}{\circle*{1}}
\multiput(105.11,115.77)(0,-15){3}{\circle*{1}}
\multiput(108.48,114.85)(0,-15){3}{\circle*{1}}
\multiput(111.86,113.92)(0,-15){3}{\circle*{1}}
\multiput(115.17,112.80)(0,-15){3}{\circle*{1}}
\multiput(118.45,111.58)(0,-15){3}{\circle*{1}}
\multiput(121.69,110.26)(0,-15){3}{\circle*{1}}
\multiput(124.90,108.85)(0,-15){3}{\circle*{1}}
\put(126,79){\vector(2,-1){20}}
\put(126,79){\vector(-1,-2){10}}
\put(149,62){\makebox(35,8)[l]{$x$}}
\put(112,49){\makebox(35,8)[l]{$z$}}
\put(126,82){\makebox(35,8)[l]{$p$}}
\put(265,120){\makebox(35,8)[l]{$f(v_z)$}}
\put(338,40){\makebox(35,8)[l]{$v_z$}}
\put(280,35){\line(0,1){75}}
\put(280.5,110){\vector(0,1){1}}
\put(250,45){\vector(1,0){83}}
\multiput(278,59)(-4,0){6}{\circle*{1.5}}
\multiput(282,59)(0,4){8}{\circle*{1.5}}
\multiput(286,87)(4,0){2}{\circle*{1.5}}
\put(294,87.5){\circle*{1.5}}
\put(298,88){\circle*{1.5}}
\put(302,89){\circle*{1.5}}
\put(306,90.5){\circle*{1.5}}
\put(309,92){\circle*{1.5}}
\put(312.5,94){\circle*{1.5}}
\end{picture}

\vskip 15pt

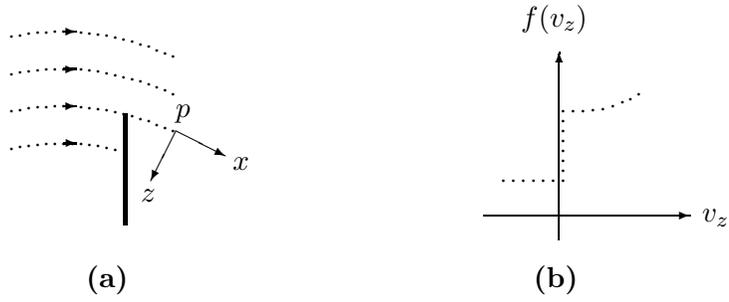
\noindent{\bf Figure 3}

\setlength{\unitlength}{0.013in} 
\begin{picture}(300,155)
\put(90,0){\makebox(35,8)[l]{\bf (a)}}
\put(270,0){\makebox(35,8)[l]{\bf (b)}}
\put(241,68){\makebox(35,8)[l]{$o$}}
\put(323,68){\makebox(35,8)[l]{$x$}}
\put(247,115){\makebox(35,8)[l]{$y$}}
\put(298,60){\makebox(35,8)[l]{\small $p$}}
\put(302,72.5){\circle*{3}}
\put(250,30){\line(0,1){80}}
\put(250.5,110){\vector(0,1){1}}
\put(310,72.5){\vector(1,0){10}}
\multiput(70,25)(10,0){2}{\line(0,1){45}}
\multiput(70,75)(10,0){2}{\line(0,1){45}}
\multiput(70,45)(0,5){12}{\line(1,0){10}}
\multiput(85,75)(170,0){2}{\vector(2,1){40}}
\multiput(85,72.5)(170,0){2}{\vector(2,0){40}}
\multiput(85,72)(170,0){2}{\line(2,-1){39}}
\multiput(124,52)(170,0){2}{\vector(2,-1){1}}
\end{picture}

\newpage
\noindent{\bf Figure 4}

\begin{picture}(200,135)
\put(90,0){\makebox(35,8)[l]{\bf (a)}}
\put(270,0){\makebox(35,8)[l]{\bf (b)}}
\put(280,35){\oval(15,10)}
\multiput(276,30)(4,0){3}{\line(0,1){4}}
\put(245,70){\vector(1,-1){5}}
\put(236,81){\line(1,-1){15}}
\put(253,27){\makebox(35,8)[l]{\small $dS$}}
\put(280,35){\vector(1,2){35}}
\put(280,35){\line(0,1){58}}
\put(280.5,93){\vector(0,1){1}}
\put(280,35){\vector(3,2){45}}
\multiput(60,70)(20,0){3}{\vector(1,-1){5}}
\multiput(51,81)(20,0){3}{\line(1,-1){15}}
\multiput(80,30)(20,0){4}{\line(4,1){20}}
\multiput(100,35)(20,0){4}{\line(0,-1){5}}
\end{picture}

\vskip 25pt

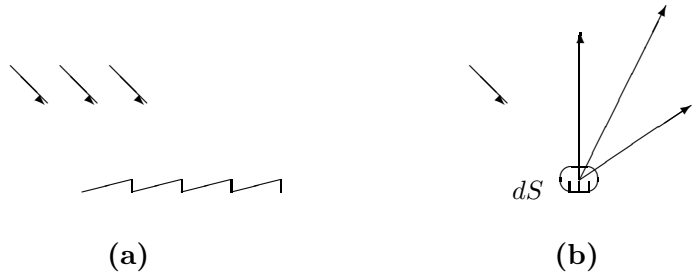
\noindent {\bf Figure 5}

\begin{picture}(200,155)
\put(90,0){\makebox(35,8)[l]{\bf (a)}}
\put(270,0){\makebox(35,8)[l]{\bf (b)}}
\multiput(85,40)(180,0){2}{\circle{30}}
\multiput(125,80)(180,0){2}{\circle*{3}}
\multiput(155,110)(180,0){2}{\circle*{3}}
\multiput(315,90)(30,30){2}{\line(1,-2){8}}
\put(343,93){\makebox(35,8)[l]{$\Delta\Omega_2$}}
\put(313,63){\makebox(35,8)[l]{$\Delta\Omega_1$}}
\multiput(112,87)(176,0){2}{\makebox(35,8)[l]{$p_1$}}
\multiput(142,113)(180,0){2}{\makebox(35,8)[l]{$p_2$}}
\put(81,40){\makebox(35,8)[l]{$o$}}
\put(90,45){\line(1,1){65}}
\put(90,45){\circle*{3}}
\multiput(85,26)(180,0){2}{\line(0,1){10}}
\multiput(90,28)(180,0){2}{\line(0,1){8}}
\multiput(95,32)(180,0){2}{\line(0,1){6}}
\multiput(80,28)(180,0){2}{\line(0,1){8}}

\multiput(75,32)(180,0){2}{\line(0,1){6}}
\multiput(305,80)(-5,-3){10}{\circle*{1}}
\multiput(305,80)(-3,-5){10}{\circle*{1}}
\put(305,80){\line(5,3){20}}
\put(305,80){\line(3,5){12}}
\multiput(335,110)(-4,-3){20}{\circle*{1}}
\multiput(335,110)(-3,-4){20}{\circle*{1}}
\put(335,110){\line(4,3){20}}
\put(335,110){\line(3,4){15}}
\end{picture}

\vskip 25pt
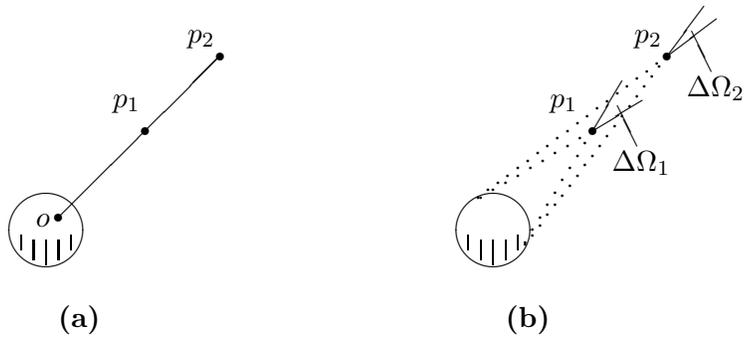
\noindent{\bf Figure 6}

\begin{picture}(200,165)

\put(90,15){\makebox(35,8)[l]{\bf (a)}}
\put(270,15){\makebox(35,8)[l]{\bf (b)}}
\multiput(52,135)(180,0){2}{\makebox(35,8)[l]{$f$}}
\multiput(158,35)(180,0){2}{\makebox(35,8)[l]{$\theta$}}
\multiput(51,40)(180,0){2}{\vector(1,0){103}}
\multiput(55.5,125)(180,0){2}{\vector(0,1){2}}
\multiput(55,35)(180,0){2}{\line(0,1){90}}
\put(100.00,115.00){\circle*{1}}
\put(103.00,114.93){\circle*{1}}
\put(97.00,114.93){\circle*{1}}
\put(106.00,114.85){\circle*{1}}
\put(94.00,114.85){\circle*{1}}
\put(108.81,113.81){\circle*{1}}
\put(91.19,113.81){\circle*{1}}
\put(111.63,112.77){\circle*{1}}
\put(88.37,112.77){\circle*{1}}
\put(113.34,110.31){\circle*{1}}
\put(86.66,110.31){\circle*{1}}
\put(115.06,107.85){\circle*{1}}
\put(84.94,107.85){\circle*{1}}
\put(116.78,105.39){\circle*{1}}
\put(83.22,105.39){\circle*{1}}
\put(117.75,102.55){\circle*{1}}
\put(82.25,102.55){\circle*{1}}
\put(118.71,99.71){\circle*{1}}
\put(81.29,99.71){\circle*{1}}
\put(119.67,96.87){\circle*{1}}
\put(80.33,96.87){\circle*{1}}
\put(120.64,94.03){\circle*{1}}
\put(79.36,94.03){\circle*{1}}
\put(121.60,91.19){\circle*{1}}
\put(78.40,91.19){\circle*{1}}
\put(122.57,88.35){\circle*{1}}
\put(77.43,88.35){\circle*{1}}
\put(123.53,85.51){\circle*{1}}
\put(76.47,85.51){\circle*{1}}
\put(124.30,82.61){\circle*{1}}
\put(75.70,82.61){\circle*{1}}
\put(125.06,79.71){\circle*{1}}
\put(74.94,79.71){\circle*{1}}
\put(125.82,76.80){\circle*{1}}
\put(74.18,76.80){\circle*{1}}
\put(126.59,73.90){\circle*{1}}
\put(73.41,73.90){\circle*{1}}
\put(127.35,71.00){\circle*{1}}
\put(72.65,71.00){\circle*{1}}
\put(128.12,68.10){\circle*{1}}
\put(71.88,68.10){\circle*{1}}
\put(128.88,65.20){\circle*{1}}
\put(71.12,65.20){\circle*{1}}
\put(129.65,62.30){\circle*{1}}
\put(70.35,62.30){\circle*{1}}
\put(132.15,60.64){\circle*{1}}
\put(67.85,60.64){\circle*{1}}
\put(135.33,61.91){\circle*{1}}
\put(64.67,61.91){\circle*{1}}
\put(280.00,115.00){\circle*{1}}
\put(283.00,114.85){\circle*{1}}
\put(277.00,114.85){\circle*{1}}
\put(285.41,113.07){\circle*{1}}
\put(274.59,113.07){\circle*{1}}
\put(286.48,110.26){\circle*{1}}
\put(273.52,110.26){\circle*{1}}
\put(287.55,107.46){\circle*{1}}
\put(272.45,107.46){\circle*{1}}
\put(288.62,104.66){\circle*{1}}
\put(271.38,104.66){\circle*{1}}
\put(289.10,101.70){\circle*{1}}
\put(270.90,101.70){\circle*{1}}
\put(289.59,98.74){\circle*{1}}
\put(270.41,98.74){\circle*{1}}
\put(290.08,95.78){\circle*{1}}
\put(269.92,95.78){\circle*{1}}
\put(290.57,92.82){\circle*{1}}
\put(269.43,92.82){\circle*{1}}
\put(291.05,89.86){\circle*{1}}
\put(268.95,89.86){\circle*{1}}
\put(291.54,86.90){\circle*{1}}
\put(268.46,86.90){\circle*{1}}
\put(291.94,83.92){\circle*{1}}
\put(268.06,83.92){\circle*{1}}
\put(292.34,80.95){\circle*{1}}
\put(267.66,80.95){\circle*{1}}
\put(292.73,77.98){\circle*{1}}
\put(267.27,77.98){\circle*{1}}
\put(293.13,75.00){\circle*{1}}
\put(266.87,75.00){\circle*{1}}
\put(293.52,72.03){\circle*{1}}
\put(266.48,72.03){\circle*{1}}
\put(293.92,69.06){\circle*{1}}
\put(266.08,69.06){\circle*{1}}
\put(294.32,66.08){\circle*{1}}
\put(265.68,66.08){\circle*{1}}
\put(294.71,63.11){\circle*{1}}
\put(265.29,63.11){\circle*{1}}
\put(296.36,60.60){\circle*{1}}
\put(263.64,60.60){\circle*{1}}
\end{picture}


\begin{thebibliography}{99}

\bibitem{reif} F. Reif, {\it Fundamentals of Statistical and Thermal 
Physics}, (McGraw-Hill book Company, 1965).
\bibitem{wirz} O.M. Belotserkovskii, {\it Computational Experiment: Direct 
Numerical Simulation of Complex Gas-dynamics Flows on the Basis of Euler,
Navier-Stokes, and Boltzmann Models}, in {\it Numerical Methods in Fluid 
Dynamicsl Physics} edited by H.J. Wirz and J.J. Smolderen, 
 p378, (Hemisphere Publishing Corporation, 1978).
\bibitem{chen0} C.Y. Chen, {\it Perturbation Methods and Statistical 
Theories}, in English, (International Academic Publishers, Beijing, 1999).  
\bibitem{harris} E.G. Harris, {\it Introduction to Modern Theoretical 
Physics}, (John Wiley and Sons, 1975).
\bibitem{kogan} M.N. Kogan, {\it Rarefied Gas Dynamics}, (Plenum Press, New 
York, 1969).
\bibitem{chen} C.Y. Chen, {\it A Path-integral Approach to the Collisionless
 Boltzmann Gas}, to be published.
\end{thebibliography}
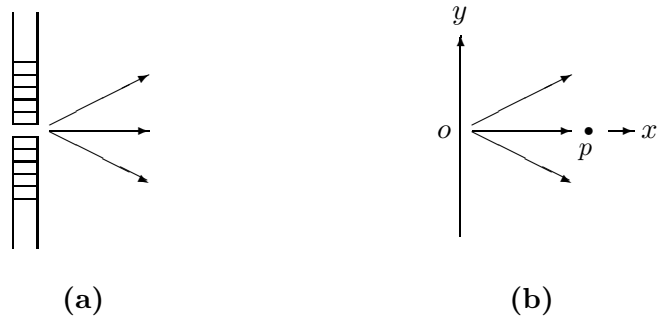
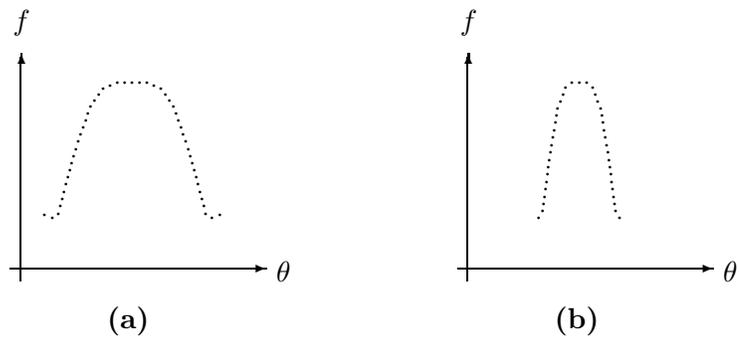
\end{document}